\documentclass[12pt]{aastex}
\usepackage{emulateapj5}
\usepackage{amssym}
\begin{document}
\title{Arcsecond Images of CH$_3$CN toward W75N}
\author{C. Watson, E. Churchwell}
\affil{Univ. of Wisconsin, Department of Astronomy, 475 N. Charter,
Madison, WI 53706}
\email{watson@astro.wisc.edu, churchwell@astro.wisc.edu}
\author{V. Pankonin}
\affil{National Science Foundation, Division of Astronomical Sciences,
4201 Wilson Boulevard, Suite 1045, Arlignton, VA 22230}
\email{vpankonin@nsf.gov}
\author{J. H. Bieging}
\affil{Steward Observatory, The University of Arizona, Tucson, AZ
85721}
\email{jbieging@as.arizona.edu}
\begin{abstract}
CH$_3$CN (J=6-5) was observed with a resolution of 2'' toward W75N
using the BIMA interferometer. Two continuum sources were detected at
3 mm, designated MM1 and MM2 in previous studies. Alignment of two mm
continuum sources with the outflow axis from MM1 suggests that these
continuum sources may be the result of the outflow interacting with
the interstellar medium. MM1 is coincident with compact CH$_3$CN
emission. CH$_3$CN was not detected toward MM2. The distribution of
optical depth ($\tau _L$) is derived. An excitation analysis was not
done because of large line optical depths.
\end{abstract}
\section{Introduction}
Hot cores in molecular clouds have a characteristic diameter of $\sim$
0.1 pc, density $\ge 10^6 cm^{-3}$ and temperature $\sim$ 100 K.
Those hot cores that are luminous in the far IR (L$_{FIR} >$ 10$^4$
L$_{\odot}$) but have no free-free radio emission are thought to be
precursors to ultra-compact (UC) HII regions (Kurtz et al, 2000 and
references therein). They are found in regions of massive star
formation, often offset from UC HII regions (Hunter et al. 2000). Hot
cores are principally studied through their molecular line emission
and mm-submm continuum emission. Hofner et al. (1996) observed the UC
HII region complex G9.62+0.19 and found that CH$_3$CN emission is
centered on the youngest, densest component in this complex. This
component also drives an energetic molecular outflow (Hofner et
al. 1996) and has weak ($<$ 1 mJy) radio continuum emission at cm
wavelengths (Testi et al. 2000). Based on observations of G9.62+0.29,
Hofner et al. (1996) postulated that CH$_3$CN emission might be a
tracer of UC HII precursors.

Wilner et al. (2001) used CH$_3$CN (J=12-11) observations of W49N to
estimate the timescale for hot core evolution; they concluded that hot
cores, which preceed the UC HII stage, have lifetimes less than or
equal to UC HII regions. Zhang et al. (1998) observed many molecular
lines, including CH$_3$CN (J=8-7), toward W51 and concluded from line
asymmetries that CH$_3$CN was tracing infalling material. These
studies found that CH$_3$CN tends to be: 1) tightly confined ($\sim$
10$^4$ - 10$^5$ AU), 2) associated with regions of massive star
formation and 3) optically thick in mm-wave transitions. Watt \& Mundy
(1999) observed CH$_3$CN (J=6-5) toward 4 massive star formation
regions (2 detections, 2 nondetections). Based on emission morphology,
excitation analysis and chemical models by Millar et al. (1997), Watt
\& Mundy (1999) concluded that the hot core near G34.26+0.15 traced by
CH$_3$CN emission was probably externally heated. In contrast to
typical results from the observations cited above, they found no
significant dust emission at 3mm. Based on single-dish detection and
interferometer non-detection, they also concluded that the CH$_3$CN
emission toward G11.94-0.62 has an extent of $\sim$ 50,000 AU.

A CH$_3$CN (J=12-11) survey was undertaken using the 10m Heinrich
Hertz Submillimeter Telescope (SMT) toward 48 known massive star
formation regions to determine the fraction with detectable CH$_3$CN
emission (Pankonin et al. 2001). Forty-six percent of the regions
surveyed were detected in CH$_3$CN emission. Because the angular
resolution was $\sim$30'', they were unable to determine if CH$_3$CN
originates from UC HII regions or from neighboring sites of even
younger precursors of UC HII regions. Pankonin et al. (2001) found
beam-averaged rotation temperatures and column densities to be
consistent with internal heating, presumably from embedded protostars
or the ionizing stars of the UC HIIs. They argued that beam dilution
allowed only lower limits for CH$_3$CN column densities.

An investigation in the lines of CH$_3$CN (J=6-5) using
interferometric resolution toward W75N is presented here to better
determine the properties of this source and to establish how important
resolution effects may be in the analysis of CH$_3$CN line
emission. W75N was selected because it is a relatively nearby region
of massive star formation region with strong CH$_3$CN emission and
there are few previous observations of the hot core in this object.

\section{Observations}
Observations of W75N were obtained with the BIMA millimeter
interferometer (Welch et al. 1996) on October 8, 11 and 12, 1999 and
April 30, 2000 in the C configuration and on October 14, 1999 in the B
configuration. Baselines ranged from 180m to 60m. The phase
calibrator, 2025+337 (J2000), was observed approximately every 25
minutes. Our primary flux density calibrator was Mars, whose
integrated flux density was 40 Jy on Oct. 10, 1999. Flux calibration
is accurate to 20\%. 2025+337 was found to have a flux density of 2.29
Jy, which is within the range of previous measurements (Welch et
al. 1996). All data editing and reduction were performed using the
MIRIAD package (Sault, Teuben, \& Wright 1995). Narrow channel
observations were obtained covering the frequency range 110.3485~GHz
to 110.3916~GHz covering the CH$_3$CN J=6-5 K=0-4 emission
lines. These windows were not wide enough to observe the nearby CH$_3
^{13}$CN transitions. Continuum subtraction was performed using two
100 MHz windows. These two windows were inspected and found to be free
of lines. All maps were made using natural weighting. The correlator
setup, rms noise and beam sizes are summarized in
Table~\ref{observation}.

\placetable{observation}
\section{Results: Continuum and CH$_3$CN Line Emission}
\subsection{MM1}
\begin{figure}
	\epsscale{.9}
	\plotone{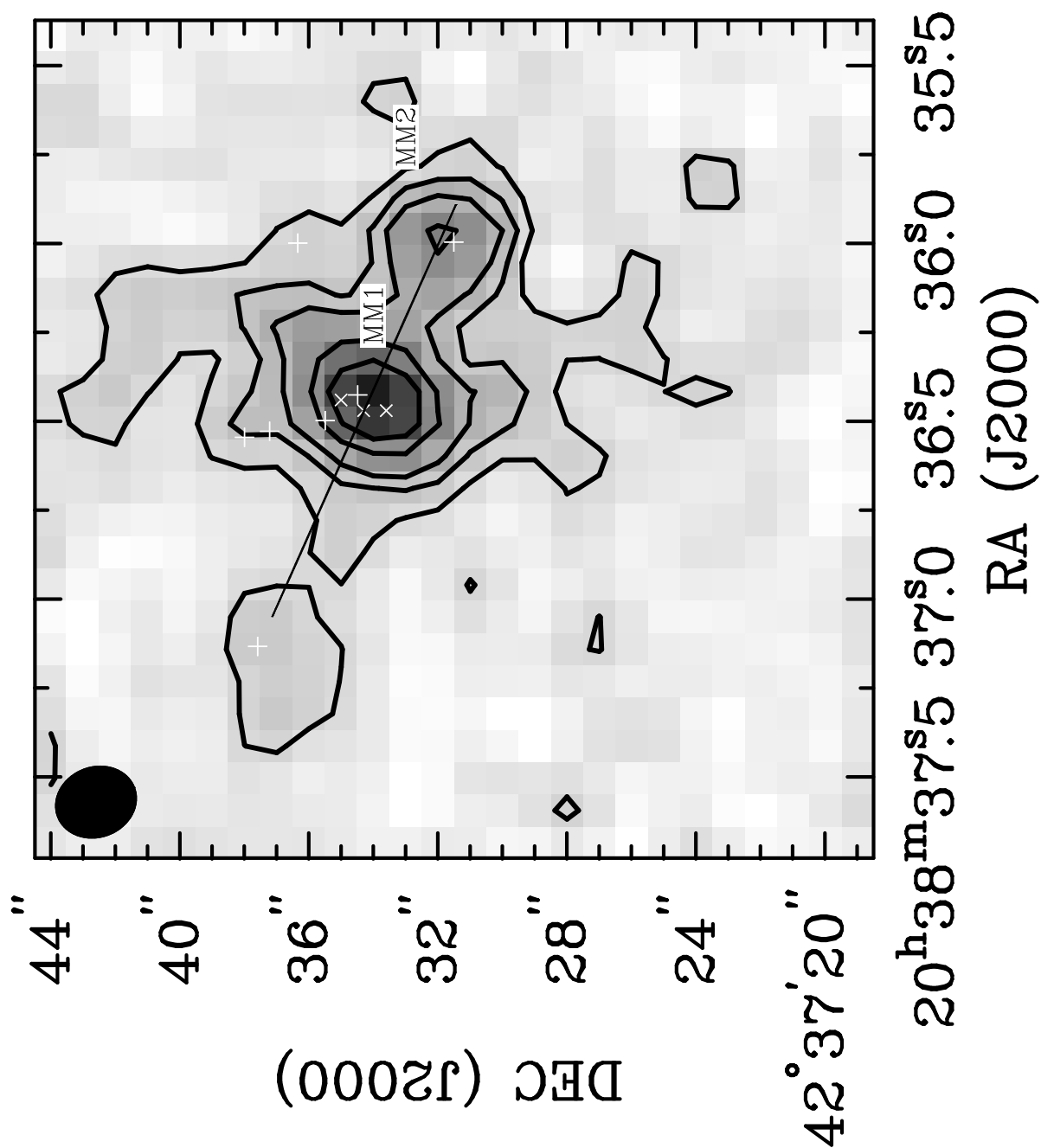}
	\caption{3mm continuum in greyscale and contours. Contours are
	at 2, 4, 6, 10 \& 14 $\sigma$ (4.2 mJy/beam). H$_2$O masers
	are represented by white plus symbols(Torrelles et
	al. 1997). VLA cm-radio continuum sources are represented by
	white Xs (Hunter et al. 1994; Torrelles et al. 1997). The CO
	and cm-radio continuum outflow orientation is indicated by a
	black line (Davis et al. 1998a; Torrelles et al. 1997). The
	synthesized HPBW is indicated in the top-left corner. The
	origin of the outflow is not well
	determined. \label{continuum}}
\end{figure}

\begin{figure}
	\plotone{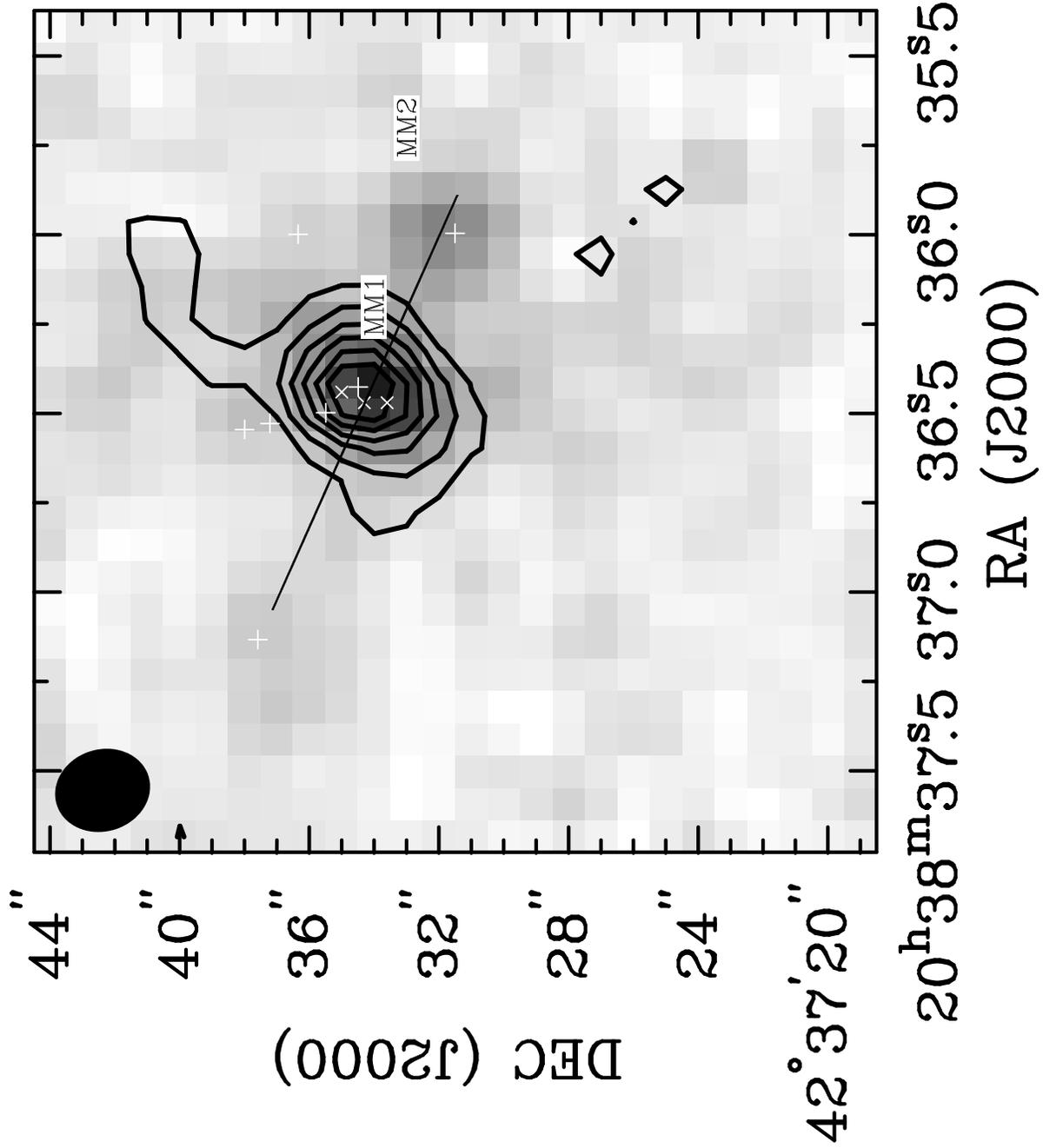} \caption{CH$_3$CN emission lines in
	contour over 110 GHz continuum in greyscale. Contours are 5,
	9, 13, 17 \& 21 $\sigma$ (8.6 mJy/beam). The CO and cm-radio
	continuum outflow orientation is indicated by a black line
	(Davis et al. 1998a; Torrelles et al. 1997). The synthesized
	HPBW is indicated in the top left corner. \label{emline}}
\end{figure}

\begin{figure}
	\plotone{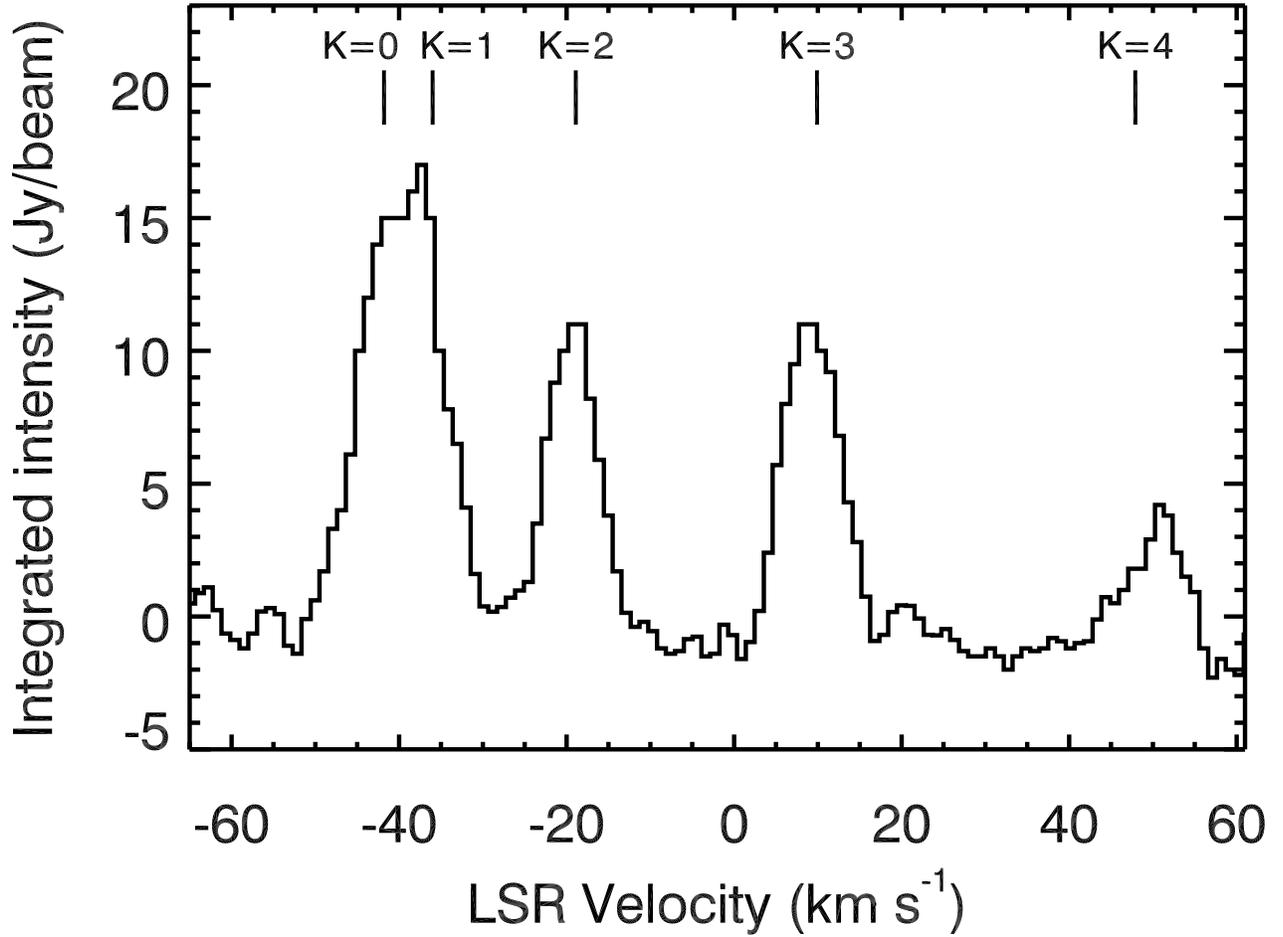} \caption{Spectrum of CH$_3$CN
	emission integrated over a 4'' by 4'' box centered on MM1. LSR
	velocity is given relative to the J=6-5 K=3 component. The
	conversion to brightness temperature is 18.2 K/(Jy/beam). No
	reasonably abundant molecular species with low excitation
	transitions were found blended with the CH$_3$CN emission
	lines. \label{pixelspectrum}}
\end{figure}

Our 3mm continuum image is shown in Figure~\ref{continuum} in both
contours and greyscale.  The line emission integrated over the K=0-4
components is shown in Figure~\ref{emline} as contours superimposed on
the 3mm continuum shown in greyscale. The CH$_3$CN (J=6-5) spectrum
integrated over a 4'' x 4'' area centered on MM1 is shown in
Figure~\ref{pixelspectrum}. The line emission is confined to MM1. As
traced by CH$_3$CN, MM1 is one of the most compact hot cores yet
observed ($\theta _{deconvolved}$ = 5.0'' x 1.5'' = 10,000 AU x 3,000
AU). The distribution of line emission is slightly elliptical, with a
deconvolved major-to-minor axis ratio of 3.4 and position angle of
24$^\circ \pm 10^\circ$ west of north, determined by a least-squares
2D Gaussian fit to the line emission. The individual K-components were
imaged separately and found to be indistinguishable from the
integrated line emission. The MM1 continuum emission is also elongated
with a position angle, 24$^\circ$ west of north but deconvolved axis
ratio of 1.6. MM1 has a peak 3mm continuum flux density of 80 $\pm$ 16
mJy/beam and integrated continuum flux density of 390 $\pm$ 80 mJy
(integrated over 200 square arcsecs). The results of our analysis are
given in Table~\ref{results}.

\placefigure{continuum}
\placefigure{emline}
\placefigure{pixelspectrum}

Previous observations with the VLA show that 3 compact cm continuum
sources lie within the central region of MM1 (Hunter et al. 1994 at
1.3 cm and 3.6cm and Torrelles et al. 1997 at 1.3 cm). Those
observations had beam sizes $\sim$ 0.1'', which permitted them to
resolve the cm-sources (see Figure~\ref{continuum}, each VLA source is
represented by a white X). The sources lie roughly along the mm
continuum major axis of MM1.  The mm continuum and CH$_3$CN major axes
are both perpendicular to the MM1 outflow axis (discussed below). This
alignment is suggestive of an oblate cloud core, although the mm and
CH$_3$CN morphology could also result from the alignment of the three
compact VLA continuum sources. MM1 and MM2 were not detected by the
MSX satellite (Mill et al. 1994) between 8$\mu$m and
21$\mu$m. Modeling the emission as thermal dust emission at T=35 K
(see section 3.3.2), we estimate that expected flux densities (S$_\nu
<$ 1 mJy) are below the MSX detection limits (S$_\nu >$ 100 mJy).

\subsection{MM2}
A second mm-continuum source, MM2, was found $\approx$ 5'' to the
southwest of MM1. MM2 has a peak 3mm continuum flux density of 45
$\pm$ 9 mJy/beam and integrated continuum flux density of 200 $\pm$ 40
mJy. MM2 was not detected at cm wavelengths, indicating that its mm
emission is probably due to dust. Previous submm observations do not
have the spatial resolution to distinguish between MM1 and MM2 (Hunter
et al. 2000).  MM2 was also not detected via K-band photometry by
Moore et al. (1988, 1991) or MSX (8$\mu$m to 21 $\mu$m).  Absence of
emission at K-band is probably due to intrinsically weak emission,
although high optical depth may also be important at $\lambda \le 2
\mu$m. There are four possible explanations for these observations
which we cannot distinguish between: 1) MM2 is heated by a proto
O-star which has not formed a detectable HII region because of infall,
2) MM2 is heated by a non-ionizing star, 3) MM2 is heated by internal
shocks, probably associated with the outflow from MM1, or 4) MM2 is
heated externally, probably by the three cm-components of MM1.

\subsection{Outflow}
A bipolar outflow has been detected toward MM1 in CO (J=3-2) (Davis et
al. 1998a; Hunter et al. 1994), CO (J=2-1) (Davis et al. 1998b) and in
cm-radio continuum (Torrelles et al. 1997); all of whom find that the
outflow position angle is 66$^\circ$ east of north (indicated by a
black line in Figures~\ref{continuum} and ~\ref{emline}). The outflow
source has not been clearly determined. The outflow axis is
perpendicular to the mm-continuum and CH$_3$CN major axis.  The cm
continuum is likely produced by free-free emission, which, assuming
the source is optically thin, would not be detectable at 3 mm. H$_2$O
and OH masers have also been measured along the outflow axis, although
no significant velocity gradient was measured (Hunter et al. 1994 and
Torrelles et al. 1997). Shepherd (2001) observed W75N at 3 mm and 1 mm
and found a third mm source (MM3) $\sim$10'' to the northeast of MM1,
hinted at by the lowest contour in Figure~\ref{continuum}. MM3, like
MM2, lies along the outflow axis but on the opposite side of MM1
relative to MM2.  Using flux density measurments at 1~mm (230 GHz) and
3~mm (90 GHz) of Shepherd (2001), the spectral indices of MM2 and MM3
are 3.1 and 1.8, respectively.  Comparing Shepherd's measurements with
those in this study, we estimate the error in the spectral index of
MM2 to be large, $\sim$1. The alignment of MM2 and MM3 suggests
that they are associated with the outflow from MM1. In this scenario
and based on its spectral index, emission from MM2 and MM3 could
originate either from shock heated dust or shock-ionized optically
thick gas in the outflow.

\placetable{results}
\subsection{Analysis}
\subsubsection{Boltzmann Analysis and Optical Depth}
CH$_3$CN is a symmetric top molecule whose K-components are closely
spaced in frequency but have a wide range of excitation energies above
ground. J=6-5, K=0 has an excitation energy of 13 K above ground,
whereas the K=3 transition, which is only 19 MHz lower in frequency,
is 76 K above ground. Assuming the line emission from MM1 is optically
thin, an excitation analysis, based on the radiation transfer
equation, allows a determination of the rotation temperature
(T$_{rot}$) and column density (N$_{tot}$).

The pattern of K-components were simultaneously fit with
Gaussian profiles. That is, we fixed the line separations and shifted
the overall pattern to obtain a least squares fit. We also forced each
line to have the same FWHM, true if all emission lines originate in
the same volume. Thus, we allowed only 6 free parameters at each
position: one common width (FWHM), one pattern velocity and 4
intensities (T$_A ^*$). The Gaussians were fit using a
Levenberg-Marquardt least-squares minimization technique (Argonne
National Labs, Minipak Project 1980). We discarded fits to spectra at
any position for which T$_A ^* < 3 \sigma$ for any single emission
line.

\begin{figure}
	\plotone{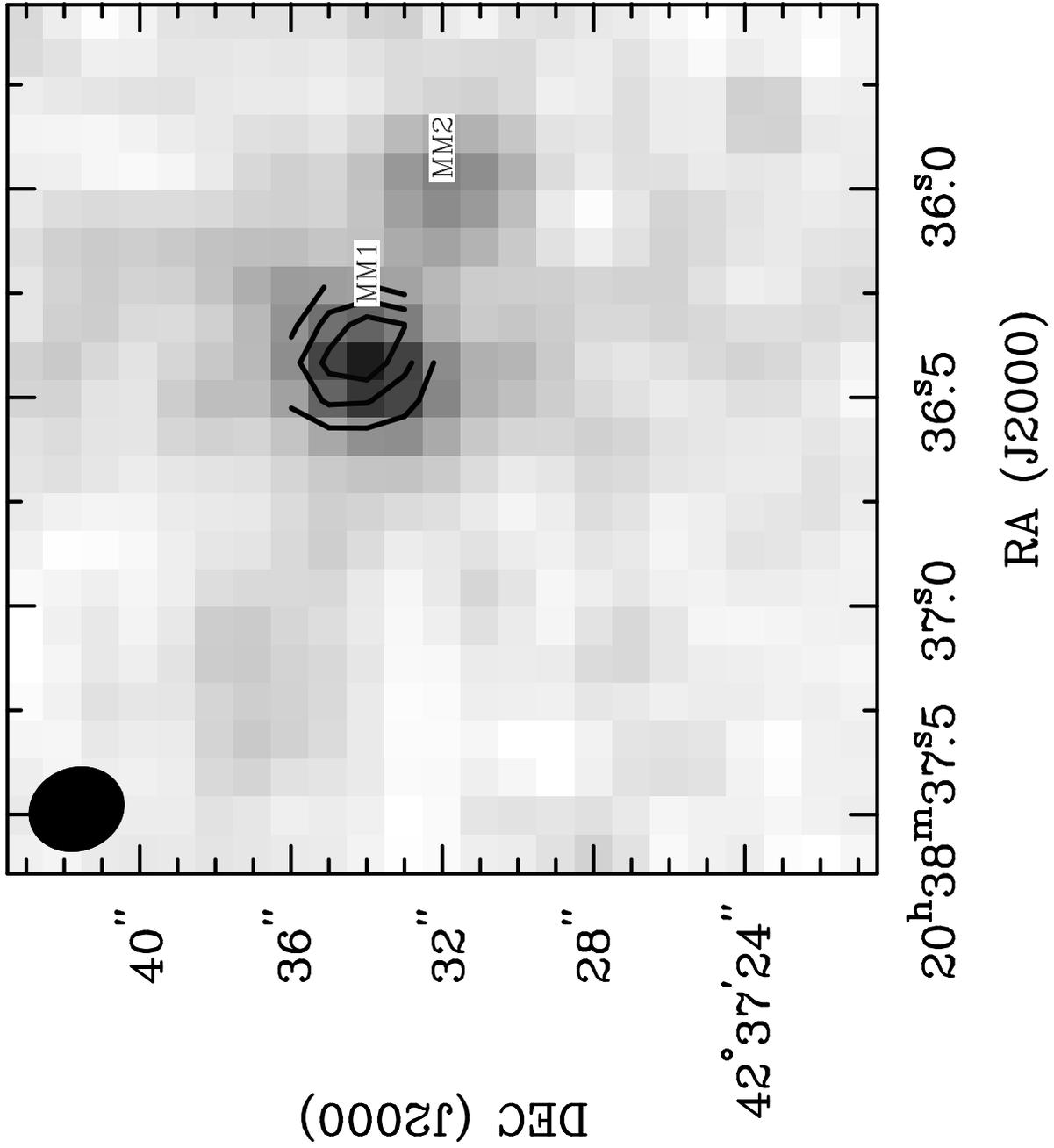} \caption{Beam averaged optical depth of the
	K=3 component in contours overlaid upon the continuum in
	greyscale. Contours are 3, 5 and 7. Maximum and minimum are
	8.5 and 0.7, respectively. The optical depth was calculated
	from the peak T $_A ^*$ ratio for K=2 and K=3
	lines. \label{tau}}
\end{figure}

Some of the assumptions on which the excitation analysis rests are not
strictly true. Specifically, the ratios of K-component intensities
show that $\tau _L \ge$ 1 in MM1 for many lines. Figure~\ref{tau}
shows the K=3 optical depth as calculated from its ratio with the K=2
line. The K=3 optical depth was found to range from 0.7 to $\sim$
8.5. If the emission lines originate from different volumes of gas,
however, this optical depth measurement would not be accurate. The
optical depth of the K=0 and K=1 lines were not calculated because of
blending. The optical depth of the K=4 line was not calculated because
the line was not strong enough for a spatially resolved
analysis. Goldsmith \& Landger (1999) studied the effects of optical
depth on the Boltzmann analysis and concluded that in such an analysis
optically thick emission lines can produce a large scatter. In order
to avoid calculating erroneous excitation temperatures and column
densities, Goldsmith \& Landger (1999) recommend combining
observations of lines with a wide range of rotational states. Pankonin
et al. (2001) compared results between their Boltzmann analysis and
three different statistical equilibrium analyses and found that the
average difference between the two methods in T$_{rot}$ was
$\sim$30\%. Pankonin et al. (2001) also derived lower limits for
N$_{tot}$ which were a factor of $\sim$3 lower than the statistical
equilibrium analyses. Because the optical depth of these lines is
large, we limit our analysis to the determination of line optical
depths from line data and mass from continuum data. Pankonin et
al. (2001) found that in a sample of 20 sources with 5 detected
CH$_3$CN (J=12-11) K-components, 8 had K=2 component optical depths
$>$ 8. Since W75N was among the 12 sources they reported without an
optical depth measurement, our analysis here implies that CH$_3$CN
line emission from hot cores frequently may be optically thick. In
such cases, a source model including kinematics coupled with a
radiative transfer/statistical equilibrium analysis is essential to
correctly determine physical properties such as T$_{rot}$ and
N$_{tot}$. Observations with significantly higher resolution
observations, high signal-to-noise and a wide range of excitation
energy will be required to provide enough constraints to justify such
a model.

The CH$_3$CN emission observations toward W75N show some similarity to
observations toward IRAS 20126+4104 (Cesaroni et al. 1999). That study
concluded that CH$_3$CN was tracing an infalling accretion disk. The
resemblance, however, is only strong enough to encourage further
observations as described above.

\placefigure{tau}
\subsubsection{Mass Estimate}
We can also approximate the mass in MM1 and MM2 using the measured
mm-continuum. We estimate the free-free emission contribution to the 3
mm emission by extrapolating from observations at 1.3 cm (Torrelles et
al. 1997) and 3.6 cm (Hunter et al. 1994). The three cm-components of
MM1 all have rising spectra from 3.6 cm to 1.3 cm, indicating the
sources are partly optically thick.  Using the spectral index between
3.6 cm and 1.3 cm to esimate the flux density at 3 mm due to free-free
emission would predict $\sim$ 100 mJy.  If the sources are optically
thin shortward of 1.3 cm, however, we estimate the free-free component
would be $\sim$20 mJy at $\lambda$ = 3 mm. Since we wish to obtain an
upper mass limit (see below) and would prefer to overestimate the dust
emission, we assume the free-free component of the 3 mm emission of
MM1 is $\sim$20 mJy. Assuming thermal dust emission accounts for the
remaining 3 mm emission (370 mJy) and following the method of
Hildebrand (1983), mm-continuum emission is related to the total mass
(gas + dust) according to:
\begin{equation}
M_{total} \ge \frac{F_\nu\,D^2}{B_\nu (T_{dust})\,\kappa_\nu}
\end{equation}
where D is the distance to the source, F$_\nu$ is the continuum flux
density, B$_\nu$ is the Planck function at frequency $\nu$ and
$\kappa_\nu$ is the opacity of dust at $\nu$. M$_{total}$ is given as
a lower limit because the measured flux density may not represent the
entire range of dust temperatures present. We assume a gas-to-dust
ratio of 100 and thus $\kappa_\nu = 0.006\biggl(\frac{\nu}{245
GHz}\biggr)^\beta$ cm$^2$ g$^{-1}$ (Carpenter 2000). We assume a
frequency index, $\beta$ = 1.5 (Shepherd 2000, Pollack et al. 1994)
for the dust opacity. This result is sensitive to the assumed value of
T$_{dust}$. If we use a lower limit for T$_{dust}$ = 10 K, however,
the relationship above gives an upper mass limit of 440 M$_\odot$ for
MM1 (see Table~\ref{results}). If more than $\sim$500 M$_\odot$ were
present at any reasonable temperature, we would measure a greater flux
density at 3 mm. Since typical values for the mass of clouds that give
rise to O-stars are $\gtrsim$ 1,000 M$_\odot$ (see Hunter et
al. 2000), we conclude that W75N is unlikely to be forming an
O-star. This result is consistent with Hunter et al. (1994) who
inferred the spectral types of the three cm sources comprising MM1 to
be early B-stars. Since Hunter et al. (1994) did not account for dust
absorption or free-free self-absorption, however, the true spectral
types may be earlier. The mass upper limit calculated for MM2 using
the same method is reported in Table~\ref{results}.

\section{Conclusions}
CH$_3$CN (J=6-5) was observed toward W75N using the BIMA
interferometer with 2'' resolution. Five compact ($\sim$ 10,000 AU)
K-components were detected (K=0-4) toward the peak of MM1 . CH$_3$CN
emission is elongated with a deconvolved major-minor axis ratio of
3.4. The major axis coincides with the alignment of three cm-continuum
sources reported previously with 0.1'' resolution VLA
observations. Thus, the elongation is probably caused by multiple
sources lying approximately along a straight line projected on the
plane of the sky, although we cannot rule out CH$_3$CN tracing out an
oblate cloud core. A bipolar outflow was previously detected toward
MM1 in CO(J=1-0, J=3-2) and cm-continuum. MM2 and MM3 (detected by
Shepherd 2001) lie on opposite sides of MM1 along the outflow axis,
indicating that they may be heated by shocks as the outflow interacts
with the local interstellar medium.

Analysis of K=2 and 3 emission lines indicates that CH$_3$CN (J=6-5)
is optically thick ($\tau _{max} \approx$ 8). Thus, a radiative
transfer/statistical equilibrium model which encorporates source
kinematics and temperature and density gradients will be required to
estimate T$_{rot}$ and N$_{CH_3CN}$. The results of this study support
the previously observed pattern that CH$_3$CN in massive star
formation regions tends to be compact ($\sim$ 10$^4$-10$^5$ AU) and
optically thick in mm-wave transitions.  

\acknowledgments We would like to thank Rick Forster and BIMA staff
for aid in reducing these data. C.W. is partially supported by the
Wisconsin Space Grant Consortium. E.B.C. acknowledges partial support
from NSF grant AST-9617686. We would like to thank an anonymous
referee whose comments helped strengthen the paper. The discussion and
conclusions in this paper do not represent the views of the NSF.

\begin{deluxetable}{lrl}
\tablecaption{Observational Summary 	\label{observation}} 
\startdata
\tableline
\tableline
Source	&W75N\\
Phase center&\\
RA (J2000)	&20$^h$ 38$^m$ 36.6$^s$\\
DEC	&42$^{\circ}$ 37' 32''\\
Transitions &CH$_3$CN (J=6-5) K=0-4\\
Center rest frequency &110.3701 GHz\\ 
Spectral resolution &1.1 km/s\\ 
Continuum Bandwidth	&200 MHz\\ 
Antenna HPBW	&96''\\
Synthesized HPBW	&2.5'' x 2.2''\\ 
Synthesized beam position angle	&21$^\circ$ E of N\\
T$_{brightness}$ conversion	&18.2 K/ (Jy/beam)\\
Max spatial scale	&60''\\
RMS$_{channel}$	&136 mJy/beam\\
RMS$_{cont}$	&4.2 mJy/beam\\ 
Flux calibrator	&\\
Mars (Oct 10 1999):	&40$\pm$8 Jy\\
Phase calibrator&\\
2025+337 (Oct 8-12, 1999)	&2.3$\pm$0.4 Jy\\
\enddata
\end{deluxetable}
\begin{deluxetable}{rrr}
\tablecaption{Continuum Results and Analysis		\label{results}}
\startdata
\tableline
\tableline
Source	&MM1	&MM2\\
RA	&20$^h 38^m 36.42^s \pm .03^s$	&20$^h 38^m 36.0^s \pm .1^s$\\
DEC	&42$^\circ 37' 34''.0 \pm 0''.5$	&42$^\circ 37' 32'' \pm 1''$\\
Peak S$_\nu$ (mJy/beam)	&80 $\pm$ 16	&45 $\pm$ 9\\
Integrated S$_\nu$ (mJy)	&390 $\pm$ 80	&200 $\pm$ 40\\
Mass upper limit (M$_\odot$)	&$<$ 440 &$<$ 240\\
\enddata
\end{deluxetable}
\end{document}